\documentclass[10pt,english,floats]{revtex4}
\usepackage[T1]{fontenc}
\usepackage[latin9]{inputenc}
\usepackage{amsmath}
\usepackage{amssymb}

\makeatletter

\usepackage{enumerate}\usepackage{xcolor}\definecolor{DarkGray}{rgb}{0.1,0.1,0.5}
\usepackage[colorlinks=true,breaklinks, linkcolor= DarkGray,citecolor= DarkGray,urlcolor= DarkGray]{hyperref}\usepackage{latexsym}\usepackage{revsymb}\usepackage{epsfig}\usepackage{pstricks}\usepackage{ifthen}\usepackage{mathrsfs}

\usepackage{amsfonts}\usepackage{subfigure}\usepackage{amsthm}\newcommand{\appref}[1]{\hyperref[#1]{{Appendix~\ref*{#1}}}}
\newcommand{\be}{\begin{eqnarray} \begin{aligned}}
\newcommand{\ee}{\end{aligned} \end{eqnarray} }
\newcommand{\benn}{\begin{eqnarray*} \begin{aligned}}
\newcommand{\eenn}{\end{aligned} \end{eqnarray*} }

\newcommand*{\cC}{\mathcal{C}}

\newcommand*{\cH}{\mathcal{H}}

\newcommand*{\cS}{\mathcal{S}}

\newcommand*{\cZ}{\mathcal{Z}}

\newcommand{\bc}{\begin{center}}
\newcommand{\ec}{\end{center}}

\newtheorem{theorem}{Theorem}[section]
\newtheorem{lemma}[theorem]{Lemma}

\newcommand{\myacknowledgments}{\begin{center}{\bf Acknowledgments}\end{center}\par}

\def\01{\{0,1\}}

\newcommand{\ket}[1]{|#1\rangle}
\newcommand{\bra}[1]{\langle#1|}

\newcommand{\ketbra}[2]{|#1\rangle\langle#2|}

\newcounter{protoCount}
\newcounter{protoList}
\newsavebox{\tmpbox}
\newlength{\protobox}

\makeatother

\usepackage{babel}

\begin{document}

\title{Classification of the phases of 1D spin chains with commuting Hamiltonians}

\author{Salman Beigi}

\email{salman.beigi@gmail.com}

\affiliation{School of Mathematics\\
 Institute for Research in Fundamental Sciences (IPM)\\
 Tehran, Iran}
\begin{abstract}
We consider the class of spin Hamiltonians on a 1D chain with periodic
boundary conditions that are (i) translational invariant, (ii) commuting
and (iii) scale invariant, where by the latter we mean that the ground
state degeneracy is independent of the system size. We correspond a directed graph to 
a Hamiltonian of this form and show that the structure of its ground space can be read from the 
cycles of the graph. We show that the
ground state degeneracy is the only parameter that distinguishes the phases
of these Hamiltonians. Our main tool in this paper is the idea of
Bravyi and Vyalyi (2005) in using the representation theory of finite
dimensional ${\text{C}}^{\ast}$-algebras to study commuting Hamiltonians.

\end{abstract}
\maketitle

\section{Introduction}\label{sec:intro}

Classification of the phases of matter is a major problem at the heart
of recent activities in quantum many-body physics research, especially
after the discovery of topologically ordered phases. The quantum double
model of Kitaev~\cite{Kitaev03} and the string-net condensation
model of Levin and Wen~\cite{Levin-Wen} show that the phase diagram
of 2D systems can be very complicated. These two models of commuting
Hamiltonians are exactly solvable and exhibit a rich structure in their
ground states as well as elementary excitations. In 1D, however, we
are not aware of any essential example of a \emph{commuting} Hamiltonian besides the Ising model. Moreover,
simulation of the ground states based on tensor networks has been
proven to be efficient in 1D gapped systems~\cite{Hastings07}.
So we expect to have an easier theory in 1D. Before reviewing known
results in this regard let us first describe what we exactly mean
by a phase.

Hamiltonians of interest are \emph{translational invariant} spin
Hamiltonians defined on a lattice with periodic boundary conditions.
We assume that Hamiltonians are \emph{scale invariant}, meaning that
the ground state degeneracy is independent of the system size, and
also have a constant gap above the ground state energy which again
is independent of the size of the system. We say that two such Hamiltonians
(their associated ground spaces) belong to the same phase if there
exists a continuous path of Hamiltonians satisfying the above conditions
starting with one of them and ending with the other.

Chen et al.~\cite{Chen11} and Schuch et al.~\cite{Schuch10}
have recently shown that assuming the ground space of such a 1D Hamiltonian
is \emph{exactly} describable by Matrix Product States (MPSs)~\cite{Fannes}, 
the only parameter that distinguishes phases is the ground
state degeneracy \footnote{1D systems with symmetries have also been studied in these papers.}. Although it seems that we can remove the assumption
in this result by using the proven area law in 1D~\cite{Hastings07},
there are some technical obstacles. First, we are able to approximate
the ground state of any gapped 1D Hamiltonian by an MPS, whose bond
dimension \emph{scales} (polynomially) with the system size. In~\cite{Chen11}
and~\cite{Schuch10} however, the bond dimension is assumed to be
fixed independent of the system size. Second, we need to show that
the true ground state and the approximate MPS are in the same phase
which again should be proved with a gap that is independent of the
system size. 

Yoshida~\cite{Yoshida11a} has considered the problem of classification
of phases in the special case of stabilizer Hamiltonians. He has shown
that any translational invariant and scale invariant stabilizer
Hamiltonian in 1D is equivalent to some independent copies of the
Ising model, and in 2D we essentially have the 2D Ising model and
the toric code. He has proved a similar result in 3D assuming some
extra conditions~\cite{Yoshida11b}. Although stabilizer Hamiltonians
seem very restrictive, they have been extensively studied in the theory
of quantum error correcting codes and quantum memories because of
their simple structure of syndrome measurements. The advantage of
these results comparing to~\cite{Chen11,Schuch10} is that Yoshida's
classification 
gives a characterization of \emph{logical operators}, and consequently low anergy excitations as well. We emphasis on the latter because
for instance in the models of quantum double and string-net condensation
the structures of elementary excitations are much richer than that
of ground states.

In this paper we generalize Yoshida's results in 1D. We consider frustration free commuting
Hamiltonians that are both translational and scale invariant. Due to the commutativity
assumption these Hamiltonians are automatically gapped. We associate a graph with a Hamiltonian
of this form and show that cycles of this graph encodes ground space of the Hamiltonian. We also comment
that the low energy excited states are described by paths of the graph. 
Our main result is that the phases of the
ground states of these systems are determined only by their degeneracy.

Although our assumption of commutativity comparing to that of~\cite{Chen11,Schuch10}  seems very restrictive, this at least
can be easily verified. Even the scale invariance of a Hamiltonians is hard to check in general, but for commuting Hamiltonians we 
provide a simple criterion to verify that. Moreover, in the last section we show that our results can be applied to a wider class of Hamiltonians than commuting ones.

\section{Some key observations}\label{sec:key}

Let us first fix some notations. Hilbert spaces are denoted
by $\cH$, and $\cH_{A}$ is the Hilbert space corresponding to \emph{register}
$A$. $\openone_{A}$ is the identity operator of $\cH_{A}$, and
$\mathbf{L}(\cH_{A})$ is the space of linear operators acting on
$\cH_{A}$. We may consider $\mathbf{L}(\cH_{A})$ as a Hilbert space
equipped with the Hilbert-Schmidt inner product.

In this paper we are interested in the phases of ground spaces of Hamiltonians define on a 1D chain with periodic boundary condisions. We say that ground spaces of two Hamiltonians belong to the same phase if there exists a continuous path of \emph{gapped} Hamiltonians starting with one of them and ending with the other. Here we often consider  translational and scale invariant Hamiltonians in which case we further assume that all Hamiltonians along the path are translational and scale invariant and their gap is independent of the system size. We do not impose this constraint for commuting Hamiltonians, i.e., two commuting Hamiltonians are said to belong to the same phase even if the path connecting them consists of non-commuting ones. This is because we are considering phases of ground spaces, and a phase is usually defined by translational and scale invariant Hamiltonians. So these two properties are essential in the definition of the problem but being commuting is an extra constraint which we impose for simplification.

By this definition when two Hamiltonians $H$ and $H'$ have the same ground spaces, they belong to the same phase. In fact, the linear interpolation $tH + (1-t)H'$, $0\leq t\leq 1$, is the path between them and one can easily show that these Hamiltonians for all $t$ are gapped. 
Local unitaries also do not change the phase of a ground space. That is Hamiltonians $H=\sum_j h_{j, j+1}$ and $H'=\sum_j Uh_{j, j+1}U^{\dagger}$, for unitary $U$ belong to the same phase, and the path $H(t) = \sum_j e^{itX} h_{j, j+1}  e^{-itX}$, $0\leq t\leq 1$, where $U=e^{iX}$, connects the two Hamiltonians. Note that translational and scale invariance are preserved in these two interpolations. We may also embed the Hilbert space of local spins into larger Hilbert spaces (and modify the local terms of the Hamiltonian). This does not change the ground space and we again obtain the same phase.

Here we focus only on \emph{frustration free} Hamiltonians. Thus by a shift of energy we may assume that the local terms of the Hamiltonian are positive semidefinite and the ground space energy is zero. Let $H=\sum_j h_{j,j+1}$ be such a Hamiltonian where $h_{j,j+1}$ acts only on sites $j$ and $j+1$. Let $P_{j,j+1}$ be the orthogonal projection onto the positive eigenspaces of $h_{j,j+1}$, i.e., $P_{j,j+1}$ is a projection and $\ker h_{j,j+1} = \ker P_{j,j+1}$. It is straightforward to see that $H$ is gapped if and only if $H'=\sum_j P_{j, j+1}$ is gapped. Moreover, $H$ and $H'$ have exactly the same ground spaces and by the above observation belong to the same phase. This means that without loss of generality we may assume that the local terms of the Hamiltonians are projections. Note that if $H$ is commuting then $H'$ is commuting as well. This is because $P_{j,j+1}$ belongs to the algebra generated by $h_{j,j+1}$ and the identity operator.

The following lemma due to Bravyi and Vyalyi~\cite{BravyiV} is a key step towards understanding the ground space of commuting Hamiltonians. For the sake of
completeness we also include a sketch of the proof. 
\begin{lemma}
\label{lem:comm} Suppose that $X_{AB}$ and $Y_{BC}$ are two hermitian
operators acting on finite dimensional Hilbert spaces $\cH_{A}\otimes\cH_{B}$
and $\cH_{B}\otimes\cH_{C}$, respectively. Assume that $X_{AB}\otimes\openone_{C}$
and $\openone_{A}\otimes Y_{BC}$ commute. Then there exists an index
set $V$ and decomposition 
\begin{align}
\cH_{B}\cong\bigoplus_{\beta\in V}\cH_{\beta_{l}}\otimes\cH_{\beta_{r}},\label{eq:decomp}\end{align}
 such that $X_{AB}$ acts nontrivially only on $\cH_{\beta_{l}}$
and not $\cH_{\beta_{r}}$ (for all $\beta\in V$), and similarly
$Y_{BC}$ acts nontrivially only on $\cH_{\beta_{r}}$. More precisely,
if we let $\Pi_{\beta}$ be the orthogonal projection onto the
subspace $\cH_{\beta_{l}}\otimes\cH_{\beta_{r}}\subseteq\cH_{B}$,
then \begin{align*}
X_{AB}=\sum_{\beta}\left(\openone_{A}\otimes\Pi_{\beta}\right)X_{AB}\left(\openone_{A}\otimes\Pi_{\beta}\right)=\sum_{\beta}R_{A\beta_{l}}\otimes\openone_{\beta_{r}},\\
Y_{BC}=\sum_{\beta}\left(\Pi_{\beta}\otimes\openone_{C}\right)Y_{BC}\left(\Pi_{\beta}\otimes\openone_{C}\right)=\sum_{\beta}\openone_{\beta_{l}}\otimes S_{\beta_{r}C},\end{align*}
 for some hermitian operators $R_{A\beta_{l}}$ and $S_{\beta_{r}C}$
acting on $\cH_{A}\otimes\cH_{\beta_{l}}$ and $\cH_{\beta_{r}}\otimes\cH_{C}$
respectively. \end{lemma} 

\noindent\emph{Sketch of proof.} Using the Schmidt
decomposition of $X_{AB}$ and $Y_{BC}$ as vectors in bipartite Hilbert
spaces $\mathbf{L}(\cH_{A})\otimes\mathbf{L}(\cH_{B})$ and $\mathbf{L}(\cH_{B})\otimes\mathbf{L}(\cH_{C})$, and writing down the constraint $[X_{AB}\otimes \openone_C , \openone_A\otimes Y_{BC}] =0$
we find that it suffices to prove the following statement.

\begin{quote}
\emph{Let $\cS$ and $\cS'$ be two subsets of hermitian
matrices acting on $\cH_{B}$ such that all operators in $\cS$ commute
with elements of $\cS'$. Then there exists a decomposition~\eqref{eq:decomp}
such that elements of $\cS$ ($\cS'$) act trivially on $\cH_{\beta_{r}}$
($\cH_{\beta_{l}}$).}
\end{quote}

So we need to prove the above statement. Let $\cZ(\cS)$ be the space of all matrices that commute with elements
of $\cS$. Since elements of $\cS$ are hermitian, $\cZ(\cS)$
is a ${\text{C}}^{\ast}$-algebra, and it contains $\cS'$. Using the representation
theory of finite dimensional ${\text{C}}^{\ast}$-algebras~\cite{Takesaki}
we find that there exists a decomposition~\eqref{eq:decomp} such
that \begin{align*}
\cZ(\cS)\cong\bigoplus_{\beta\in V}\openone_{\beta_{l}}\otimes\mathbf{L}(\cH_{\beta_{r}}).\end{align*}
 Thus elements of $\cS'$ have the desired form. Furthermore, it is
not hard to see that a matrix commuting with such $\cZ(\cS)$ has
to act trivially on $\cH_{\beta_{r}}$. \hfill{}{$\Box$}

\section{Graphs encode ground states of 1D commuting Hamiltonians}

Consider a 1D chain of $N$ spins of finite dimension $d$ (qudits) with
periodic boundary conditions. We let the Hamiltonian be $H_{N}=\sum_{j=1}^{N}P_{j,j+1}$,
where $P_{j,j+1}$ acts only on sites $j,j+1$ (Hereafter we adopt indices to be modulo $N$). By the observation of the previous section, since we let $H_N$ to be \emph{frustration free}, we further assume that $P_{j,j+1}$'s are projections. The extra assumptions that we impose in this section are that $H_N$ is \emph{translational invariant}, and that $H_N$ is \emph{commuting}.

According to Lemma~\ref{lem:comm} since $P_{j-1,j}$ and $P_{j,j+1}$
commute, there exists a decomposition 
\begin{align}
\mathbb{C}^{d}\cong\bigoplus_{\alpha\in V}\cH_{\alpha_{l}^{j}}\otimes\cH_{\alpha_{r}^{j}}\label{eq:decomp-qudit}
\end{align}
 of the Hilbert space of the $j$-th qudit such that $P_{j-1,j}$
($P_{j,j+1}$) acts trivially on $\cH_{\alpha_{r}^{j}}$ ($\cH_{\alpha_{l}^{j}}$).
Using the fact that $P_{j,j+1}$'s are translations of each other,
we may consider the same decomposition at site $j+1$. As a result,
$P_{j,j+1}$ acts trivially on $\cH_{\alpha_{r}^{j+1}}$ as well.
Therefore, $P_{j,j+1}$ can be written as 
\begin{align}
P_{j,j+1}=\sum_{\alpha,\beta\in V}\,\openone_{\alpha_{l}^{j}}\otimes Q_{\alpha_{r}^{j},\beta_{l}^{j+1}}\otimes\openone_{\beta_{r}^{j+1}},
\label{eq:dc1}
\end{align}
where $Q_{\alpha_r^j, \beta_{l}^{j+1}}$ acts on $\cH_{\alpha_r^j}\otimes \cH_{\beta_l^{j+1}}$ and is a projection.
 Since $P_{j,j+1}$'s are shifts of each other we may ignore the index
$j$ and represent $Q_{\alpha_{r}^{j},\beta_{l}^{j+1}}$ by $Q_{\alpha_{r},\beta_{l}}$.
The combined Hilbert space of all $N$ qudits then can be written
as \begin{align}
\bigoplus_{\alpha^{1},\dots,\alpha^{N}\in V}\,\,\bigotimes_{j=1}^N\cH_{\alpha_{r}^{j}}\otimes\cH_{\alpha_{l}^{j+1}}.\label{eq:dc2}\end{align}

We correspond a \emph{directed} graph $G$ to our Hamiltonian. The
vertex set of $G$ is the index set $V$ of decomposition~\eqref{eq:decomp-qudit}.
A directed edge is drawn from $\alpha\in V$ to $\beta\in V$
if $\ker Q_{\alpha_{r},\beta_{l}}\subseteq\cH_{\alpha_{r}}\otimes\cH_{\beta_{l}}$
is non-zero. Note that this graph may contain loops, i.e., edges from a vertex to itself. Moreover, $G$
is defined based on the local terms $P_{j,j+1}$ and is independent of
the system size $N$. 

$G$ encodes the ground space of the Hamiltonian
as follows. Let $(\alpha^{1},\dots,\alpha^{N})$ be a directed cycle of length
$N$ in $G$ which may contain a vertex or edge several times. Since $\alpha^j\rightarrow \alpha^{j+1}$ is an edge of $G$, there is a non-zero vector $\ket{\varphi_{j}}$ in $\ker Q_{\alpha_{r}^{j},\alpha_{l}^{j+1}}$ for 
$j=1,\dots,N$. By decomposition~\eqref{eq:dc2}, $\vert \varphi\rangle = \ket{\varphi_{1}}\otimes\cdots\otimes\ket{\varphi_{N}}$ is a state in the Hilbert space of $N$ spins. Then using~\eqref{eq:dc1} we have
\begin{align*}
H_N \vert \varphi\rangle & = \sum_j  P_{j,j+1} \ket{\varphi_{1}}\otimes\cdots\otimes\ket{\varphi_{N}} \\
&= \sum_j \ket{\varphi_{1}}\otimes \cdots    \otimes Q_{\alpha_{r}^{j},\alpha_{l}^{j+1}} \vert \varphi_j\rangle  \otimes      \cdots \otimes\ket{\varphi_{N}}\\
&=0.
\end{align*}
So $\ket{\varphi_{1}}\otimes\cdots\otimes\ket{\varphi_{N}}$
is in the ground space of $H_{N}$. Indeed, all ground states of $H_N$ are of this form. From \eqref{eq:dc1} and \eqref{eq:dc2} we have 
\begin{align*}
\ker H_N = \bigoplus_{\alpha^{1},\dots,\alpha^{N}\in V}\,\,\,  \bigotimes_{j=1}^N   \ker  Q_{\alpha_{r}^{j},\alpha_{l}^{j+1}}.
\end{align*}
$\ker  Q_{\alpha_{r}^{j},\alpha_{l}^{j+1}}$ is non-zero iff $\alpha^j \rightarrow\alpha^{j+1}$ is an edge of $G$. Thus if we let $\cC_{N}$ to be the set of \emph{ordered} cycles of length $N$ in $G$, we have
\begin{align*}
\ker H_N = \bigoplus_{(\alpha^{1},\dots,\alpha^{N})\in \cC_N}\,\,\,  \bigotimes_{j=1}^N   \ker  Q_{\alpha_{r}^{j},\alpha_{l}^{j+1}},
\end{align*}
and
\begin{align*}
\dim\ker H_{N}=\sum_{(\alpha^{1},\dots,\alpha^{N})\in\cC_{N}}\,\,\prod_{j=1}^{N}\dim\ker Q_{\alpha_{r}^{j},\alpha_{l}^{j+1}}.
\end{align*}
Here we must consider ordered cycles because we have the freedom
of choosing which vertex corresponds to the first site, so for example
in $\cC_{2}$, $(\alpha,\beta)$ and $(\beta,\alpha)$ are counted
as different cycles if $\alpha\neq\beta$.

As a summary we conclude that the ground space of $H_N$, for all $N$, is completely characterized by the graph $G$ and the subspaces $\ker Q_{\alpha_{r},\beta_{l}}$ corresponding to each edge $\alpha\rightarrow\beta$. Similarly to the ground states that correspond to cycles of $G$, it is easy to see that excited states with energy 1 correspond to paths of $G$, and eigenstates with eigenvalue 2 correspond to unions of two paths.

Note that the translational invariance assumption in this result is not crucial. If $H_N$ is not translational invariant, instead of a single index set $V$ we have $N$ index sets $V_1,\dots, V_N$; $V_j$ corresponds to the $j$-th local term. Then the associanted graph would be $N$-partite where there are directed edges from vertices in $V_j$ to vertices in $V_{j+1}$.

\begin{figure}
\includegraphics[width=2.5in]{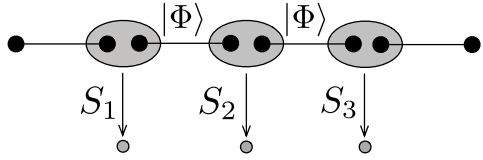}
\caption{ An $N$-qudit matrix product state (with periodic boundary conditions and) with bond dimension $\chi$ is obtained by a chain of $N$ maximally entangled states $\ket{\Phi}$ of local dimension $\chi$ followed by local maps $S_j: \mathbb{C}^{\chi}\otimes \mathbb{C}^{\chi} \rightarrow \mathbb{C}^d$. Then the final state equals $\left(S_1\otimes \cdots \otimes S_N\right)  \ket{\Phi}\otimes \cdots \otimes \ket{\Phi}$. For translational invariant MPSs we have $S_1=S_2=\dots=S_N$.
From this representation it is clear that if $\ket{\varphi}$ is an MPS, then $\left(X_1\otimes \cdots \otimes X_N\right) \ket{\varphi}$ is also an MPS with the same bond dimension. Moreover, the maximally entangled states $\ket{\Phi}$ can be replaced with any bipartite state with local dimension $\chi$ and we still obtain an MPS with bond dimension $\chi$.
}\label{fig:mps}
\end{figure}

An implication of our result is that the ground states of commuting Hamiltonians can be represented by MPSs. An MPS representation consists of a chain of maximally entangled states on which we apply some local linear transformations \cite{CiracMps} (see Figure~\ref{fig:mps}). To write the state $\ket \varphi= \ket{\varphi_{1}}\otimes\cdots\otimes\ket{\varphi_{N}}$ defined above as an MPS, we need only to map the $j$-th maximally entangled state to $\ket{\varphi_j}$. Thus $\ket\varphi$ is an MPS whose bond dimension is a constant independent of $N$.

\begin{figure}
\includegraphics[width=1.5in]{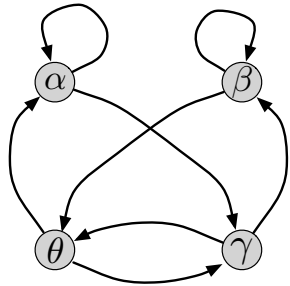}
\caption{ The graph associated with the local projection $P_{j, j+1}  = \big(\openone_j \otimes \openone_{j+1} - (\sigma_x\otimes \sigma_x)_j \otimes (\sigma_z\otimes \sigma_z)_{j+1}\big)/2$, and Hamiltonian $H_N=\sum_j P_{j, j+1}$. This Hamiltonian is commuting and translational invariant but not scale invariant.
}\label{fig:graph}
\end{figure}

Let us work out an example to clarify the ideas. Let $d=4$, and using $\mathbb{C}^4\cong \mathbb{C}^2\otimes \mathbb{C}^2$ represent each site with two spin-half particles. Define 
$$P_{j, j+1}  = \big(\openone_j \otimes \openone_{j+1} - (\sigma_x\otimes \sigma_x)_j \otimes (\sigma_z\otimes \sigma_z)_{j+1}\big)/2,$$ 
where $\sigma_x$ and $\sigma_z$ are Pauli matrices. Then $H_N=\sum_j P_{j, j+1}$ is commuting and translational invariant. The index set $V$ associated with $H_N$ has four elements and components of decomposition~\eqref{eq:decomp-qudit} are 
\begin{align*}
\cH_{\alpha_l}\otimes \cH_{\alpha_r} &= \mathbb{C}\left(  \ket{00} + \ket{11}   \right),\\
\cH_{\beta_l}\otimes \cH_{\beta_r} &= \mathbb{C}\left(  \ket{01} - \ket{10}   \right),\\
\cH_{\gamma_l}\otimes \cH_{\gamma_r} &= \mathbb{C}\left(  \ket{00} - \ket{11}   \right),\\
\cH_{\theta_l}\otimes \cH_{\theta_r} &= \mathbb{C}\left(  \ket{01} + \ket{10}   \right).
\end{align*}
Note that in this case since these subspaces are all one dimensional, we do not need to specify subsystems $\cH_{\alpha_l}$, $\cH_{\alpha_r}, \dots$ individually. The corresponding graph $G$ has four vertices and is depicted in Figure~\ref{fig:graph}. For $N=3$ for instance, $(\alpha, \gamma, \theta)$ is a cycle of length three and then $\left(  \ket{00} + \ket{11}   \right)\otimes \left(  \ket{00} - \ket{11}   \right)\otimes \left(  \ket{01} + \ket{10}   \right)$ is a ground state. However, $(\alpha, \gamma, \beta)$ is a path which cannot be completed to a cycle and then $\left(  \ket{00} + \ket{11}   \right)\otimes \left(  \ket{00} - \ket{11}   \right)\otimes \left(  \ket{01} - \ket{10}   \right)$ is an excited state with energy 1.

\section{Scale invariance}
\label{sec:scale}

We now apply the assumption that the system is scale invariant, meaning
that for sufficiently large $N$ the ground state degeneracy is a
constant independent of $N$. So we let $H_N$ be commuting, and translational and scale invariant. Our goal in this section is to show
that for every $N$ \begin{align}
\dim\ker H_{N}=\vert\cC_{1}\vert,
\label{eq:deg-loops}
\end{align}
 where $\cC_{1}$ is the set of loops of $G$.

Let $D_{1}\in\cC_{N}$ be a cycle of $G$. By going around $D_{1}$
many times, we may assume with no loss of generality that $N$, the
length of this cycle, is sufficiently large. Let $\cC_{N}=\{D_{1},\dots,D_{k}\}$
and $\cC_{N+1}=\{E_{1},\dots,E_{k'}\}$. We have 
\begin{align}
\dim\ker H_{N} & =\sum_{i=1}^{k}\,\,\prod_{\alpha\rightarrow\beta\in D_{i}}\dim\ker Q_{\alpha_{r},\beta_{l}},\label{eq:comparing1}\\
\dim\ker H_{N+1} & =\sum_{i=1}^{k'}\,\,\prod_{\alpha\rightarrow\beta\in E_{i}}\dim\ker Q_{\alpha_{r},\beta_{l}},
\end{align}
 where by $\alpha\rightarrow\beta\in D_{i}$ we mean that $\alpha\rightarrow\beta$
is an edge in the cycle $D_{i}$, and similarly for $\alpha\rightarrow\beta\in E_i$. For a cycle $D$ of length $N$
let $D^{t}$ be the cycle of length $tN$ which starting from the
same vertex as $D$, goes $t$ times around it. Then $D_{1}^{N+1},\dots,D_{k}^{N+1}$
are different cycles in $\cC_{N(N+1)}$, and similarly $E_{1}^{N},\dots,E_{k'}^{N}\in\cC_{N(N+1)}$
are distinct. Therefore, 
\begin{align}
\dim\ker H_{N(N+1)} & \geq\sum_{i=1}^{k}\,\,\prod_{\alpha\rightarrow\beta\in D_{i}}\left(\dim\ker Q_{\alpha_{r},\beta_{l}}\right)^{N+1},\\
\dim\ker H_{N(N+1)} & \geq\sum_{i=1}^{k'}\,\,\prod_{\alpha\rightarrow\beta\in E_{i}}\left(\dim\ker Q_{\alpha_{r},\beta_{l}}\right)^{N}.\label{eq:comparing4}
\end{align}
 Now applying the scale invariance assumption we have $\dim\ker H_{N(N+1)}=\dim\ker H_{N}=\dim\ker H_{N+1}$.
So by comparing equations~\eqref{eq:comparing1}-\eqref{eq:comparing4} we obtain \begin{align}
\dim\ker Q_{\alpha_{r},\beta_{l}}=1\label{eq:dim-ker-q}\end{align}
 for every edge $\alpha\rightarrow\beta$ in one of the cycles $D_{i}$
or $E_{i}$. (Note that when $\alpha\rightarrow\beta$ is an edge
of $G$, $\dim\ker Q_{\alpha_{r},\beta_{l}}$ is non-zero.) We further
conclude that $k=k'=\dim\ker H_{N(N+1)}$ and indeed \begin{align*}
\cC_{N(N+1)}=\{D_{1}^{N+1},\dots,D_{k}^{N+1}\}=\{E_{1}^{N},\dots,E_{k}^{N}\}.\end{align*}
 Thus for some $i$, $D_{1}^{N+1}=E_{i}^{N}$. By the fact that $N$
and $N+1$ are relatively prime, we find that the arbitrarily chosen cycle $D_{1}$ consists of
the repetition of a single loop. We conclude that the scale invariance assumption implies that 
all cycles of $G$ are essentially loops. Then~\eqref{eq:deg-loops} is
an immediate consequence of~\eqref{eq:dim-ker-q}.

The structure of the ground space of $H_N$ in even simpler with the scale invariance assumption. Let $\alpha\rightarrow\alpha$ be a loop of $G$, and let $\ket{\varphi_{\alpha}}\in\cH_{\alpha_{r}}\otimes\cH_{\alpha_{l}}$
be a vector that spans $\ker Q_{\alpha_{r},\alpha_{l}}$. Then $\ket{\varphi_{\alpha}}^{\otimes N}$
is a ground state, and the ground space of $H_{N}$ is spanned by these vectors
for all loops $\alpha\rightarrow\alpha$
of $G$. The vectors $\ket{\varphi_{\alpha}}^{\otimes N}$ are still MPSs and in fact translational invariant MPSs.

The example of Figure~\ref{fig:graph} is not scale invariant because the graph contains cycles that do not come from loops.
For the special case of the 1D Ising model, the graph $G$ consists
of two vertices with two loops and no other edges. From these two
loops and the above construction we obtain the all spin-up and all
spin-down states as the ground states of the Ising model.

Elementary excitations are also easier to characterize with the scale invariance assumption. As mentioned in the previous section, elementary excitations correspond to paths of $G$, and excitations with energy 2 come from unions of two paths. A sufficiently long path in a finite graph must contain a cycle. Since loops are the only cycles of such $G$, any elementary excitation of $H_N$, for sufficiently large $N$, contains several copies of states $\ket {\varphi_{\alpha}}$, for loops $\alpha\rightarrow \alpha$, in their subsystems. In fact an energy 1 eigenstate corresponds to a path of the form either $\alpha\rightarrow \alpha \cdots\rightarrow \alpha \rightarrow \beta$ or  $\beta\rightarrow \alpha\rightarrow \alpha \cdots\rightarrow \alpha $. For example, Ising model does not have any energy 1 eigenstate because its corresponding graph is a union of two loops and there is no edge of the form $\alpha\rightarrow \beta$ for $\alpha\neq \beta$.

\section{Phases are distinguished by the ground state degeneracy}\label{sec:distinguish}

We are now ready to study phases of 1D spin chains with commuting Hamiltonians. Objects of interest are ground spaces of 1D commuting Hamiltonians that are both translational and scale invariant. By the discussion of Section~\ref{sec:key} without loss of generality we can also assume that the Hamiltonian is a summation of local projections. Thus we may use results of the previous two sections which are summarized as
\begin{align}\label{eq:ground-10}
\ker H_N = \bigoplus_{\alpha\rightarrow \alpha\in \cC_1}  \left(\ker Q_{\alpha_r, \alpha_l}\right)^{\otimes N},
\end{align}
where $\ker Q_{\alpha_r, \alpha_l}$ is one dimensional and is spanned by $\vert \varphi_{\alpha}\rangle$. 

Here since we are interested only in the ground spaces we can replace the other $Q_{\alpha_r, \beta_l}$'s that do not appear in the above expression with the identity operator. More precisely, for every $\alpha\neq \beta$, or $\alpha=\beta$ where $\alpha\rightarrow \alpha$ is not a loop, define $\widetilde Q_{\alpha_r, \beta_l} = \openone_{\alpha_r}\otimes \openone_{\beta_l}$, and for loops $\alpha\rightarrow \alpha \in \cC_1$ let $\widetilde Q_{\alpha_r, \alpha_l} = Q_{\alpha_r, \alpha_l}$. Furthermore, define 
$$\widetilde P_{j, j+1} = \sum_{\alpha, \beta\in V}    \openone_{\alpha_l^j}\otimes \widetilde Q_{\alpha_r^j, \beta_l^{j+1}}\otimes \openone_{\beta_r^{j+1}},$$
and $\widetilde H_N = \sum_j \widetilde P_{j,j+1}$. Then $\widetilde H_N$ is still commuting and translational invariant. Moreover, its corresponding graph consists of the same loops as $G$, but no other edges. In fact, using \eqref{eq:ground-10}, $\widetilde H_N$ has the same ground space as $H_N$ and they are in the same phase. So we may replace $H_N$ with $\widetilde H_N$, i.e., we assume that $\widetilde P_{j,j+1} = P_{j,j+1}$ and $H_N = \widetilde H_N$.

For every loop $\alpha \rightarrow \alpha$ fix arbitrary states $\ket{\xi_{\alpha_r}} \in \cH_{\alpha_r}$ and $\ket{\xi_{\alpha_l}} \in \cH_{\alpha_l}$. Let $U_{\alpha}$ be a unitary operator that acts on $\cH_{\alpha_r}\otimes \cH_{\alpha_l}$ in such a way that $U_{\alpha} \ket {\varphi_{\alpha}}  = \ket {\xi_{\alpha_r}} \otimes \ket {\xi_{\alpha_l}}$. Recall that the Hilbert space of two qudits $j$ and $j+1$ can be decomposed as 
\begin{align}\label{eq:decomp-112}
\bigoplus_{\beta, \gamma\in V}\,\, \cH_{\beta_l^j}\otimes \cH_{\beta_r^j}\otimes \cH_{\gamma_l^{j+1}}\otimes \cH_{\gamma_r^{j+1}}.
\end{align}
Thus we can define the two-qudit unitary operator $U$ that acts as $U_{\alpha}$ on the subsystem/subspace $\cH_{\alpha_r^j}\otimes \cH_{\alpha_l^{j+1}}$ when $\alpha\rightarrow\alpha$ is a loop, and acts as the identity operator elsewhere.

Define $P'_{j,j+1}  = UP_{j,j+1} U^{\dagger} $, and $H'_N = \sum_j P'_{j,j+1}$. Since $U$ is block-diagonal with respect to the decomposition~\eqref{eq:decomp-112} and acts trivially on $\cH_{\beta_l^j}$ and $\cH_{\gamma_r^{j+1}}$, the new Hamiltonian $H'_N$ is commuting. The corresponding graph of $H'_N$ is the same as that of $H_N$ and consists of a union of loops. The only difference is that the state corresponding to the loop $\alpha\rightarrow \alpha$, is equal to $U \ket {\varphi_{\alpha}}  = \ket {\xi_{\alpha_r}} \otimes \ket {\xi_{\alpha_l}}$. 

Two Hamiltonians $H_N$ and $H'_{N}$ belong to the same phase because they differ only by local unitaries (see Section~\ref{sec:key}). So again without loss of generality we assume $P_{j,j+1}=P'_{j,j+1}$ and $H_N = H'_N$. In this step we turned $\ket{\varphi_{\alpha}}$, which belongs to $\cH_{\alpha_r}\otimes \cH_{\alpha_l}$, into a product state.

By~\eqref{eq:ground-10} the ground space of $H_N$ is spanned by vectors 
$$\bigotimes_{j=1}^N \ket{\xi_{\alpha_r^j}}\otimes\ket{\xi_{\alpha_l^{j+1}}} = \bigotimes_{j=1}^N  \ket{\xi_{\alpha_l^j}} \otimes \ket{\xi_{\alpha_r^j}} = \ket{\alpha}_1\otimes \cdots \otimes \ket{\alpha}_N,$$
for loops $\alpha\rightarrow \alpha$, where we set $\ket{\alpha}_j = \ket{\xi_{\alpha_l^j}} \otimes \ket{\xi_{\alpha_r^j}}$. If we let $\hat{H}_N =\sum_j \hat{P}_{j, j+1}$ such that 
$$\hat{P}_{j, j+1} = \openone_j\otimes \openone_{j+1} - \sum_{ \alpha\rightarrow\alpha\in \cC_1} \ket{\alpha}\bra{\alpha}_j \otimes \ket{\alpha}\bra{\alpha}_{j+1},$$
then $\hat{H}_N$ is commuting and translational invariant, and has the same ground space as $H_N$. Then they belong to the same phase. Moreover, since $\ket{\xi_{\alpha_r}}$ and $\ket{\xi_{\alpha_l}}$ were arbitrarily chosen, the only parameter that determines $\hat{H}_N$ is the size of $\cC_1$, i.e., the ground space degeneracy. For example for two loops we obtain the Ising model whose local projections are given by 
$$P = \openone\otimes \openone    - \left(   \ketbra{0}{0} \otimes \ketbra{0}{0}   +   \ketbra{1}{1} \otimes \ketbra{1}{1}        \right). $$
We conclude that the phases of ground spaces of translational and scale invariant commuting Hamiltonians in 1D are characterized by their degeneracy.

\section{Summary and Outlook}\label{sec:summary}

In this paper we described the structure of the ground states of translational
and scale invariant, 1D commuting Hamiltonians. We associated a graph with a commuting Hamiltonian which encodes the ground space in its cycles and the low energy states in its paths. Our results generalize
Yoshida's work in 1D who considers stabilizer Hamiltonians~\cite{Yoshida11a}. Comparing
to~\cite{Chen11, Schuch10} instead of assuming that
the ground states are described by MPSs, we imposed the assumption that
the Hamiltonian is commuting. 

In Section~\ref{sec:scale} we argued that ground states of the Hamiltonians under consideration 
can be exactly written as translational invariant MPSs. Thus we could have skipped the previous section and directly used the result of~\cite{Chen11, Schuch10} 
to conclude that the ground state degeneracy is the only parameter that distinguishes phases. Our arguments, however, are based on much simpler techniques and we preferred not to refer to~\cite{Chen11, Schuch10}.

Our results can be applied on a larger class of Hamiltonians than the commuting ones. Let $H_N=\sum_j h_{j,j+1}$ be an arbitrary translational invariant (frustration free) Hamiltonian. As before without loss of generality we may assume that $h_{j, j+1}$ is positive semidefinite and the ground state energy of $H_N$ is zero. Suppose that there exists a \emph{positive
definite} matrix $X$ such that 
\begin{align}\label{eq:x}
(h_{j-1,j}\otimes\openone_{j+1})\cdot(\openone_{j-1}\otimes X_{j}\otimes\openone_{j+1})\cdot(\openone_{j-1}\otimes h_{j,j+1})=(\openone_{j-1}\otimes h_{j,j+1})\cdot(\openone_{j-1}\otimes X_{j}\otimes\openone_{j+1})\cdot(h_{j-1,j}\otimes\openone_{j+1}).
\end{align}
Then the Hamiltonian $H'_N=\sum_j h'_{j,j+1}$ with local term
\begin{align}
h'_{j,j+1}=(X_{j}^{1/2}\otimes X_{j+1}^{1/2})h_{j,j+1}(X_{j}^{1/2}\otimes X_{j+1}^{1/2})
\label{eq:localterm}
\end{align}
is commuting. These local terms are still positive semidefinite, and one can easily observe that $\ket{\varphi}$ is a zero energy state of $H'_N$ if and only if $X_1^{1/2}\otimes \cdots \otimes X_N^{1/2} \ket{\varphi}$ is the ground state of $H_N$. Thus the new Hamiltonian is frustration free as well. Furthermore, since $X$ is assumed to be positive definite and then invertible, the correspondence between ground states is one-to-one. Therefore, $H'_N$ is scale invariant if and only if $H_N$ is scale invariant, and in this case results of our paper are applied. For instance, we showed that ground states of commuting Hamiltonians have MPS representations. Moreover, if $\ket{\varphi}$ is an MPS then $X_1^{1/2}\otimes \cdots \otimes X_N^{1/2} \ket{\varphi}$ is an MPS as well (see Figure~\ref{fig:mps}). As a summary, the existence of a positive definite matrix $X$ satisfying \eqref{eq:x} implies that the ground states of $H_N=\sum_j h_{j, j+1}$ have MPS representations, and results of \cite{Chen11, Schuch10} are applied.

With the above technique sometimes we can turn a Hamiltonian whose ground space has MPS representation into a commuting one. In particular the converse of this observation holds for Hamiltonians with a \emph{unique injective} MPS ground state in the following sense. Consider a translational invariant Hamiltonian with a unique translational invariant MPS ground state $\ket{\varphi}$. We assume that the bond dimension of this MPS representation is $\chi$ and the corresponding map is $S: \mathbb{C}^{\chi}\otimes \mathbb{C}^{\chi} \rightarrow \mathbb{C}^d$. Then we have 
$$\ket{\varphi}  = \left(S\otimes \cdots \otimes S\right)  \ket{\Phi}\otimes \cdots \otimes \ket{\Phi},$$  
where $\ket{\Phi}$ is the maximally entangled state with local dimension $\chi$. We may also replace $S$ with $US$ where $U$ is a unitary map. In this case $\ket{\varphi}$ is replaced with $U\otimes \cdots \otimes U\ket{\varphi}$ which by the discussion of Section~\ref{sec:key} belongs to the same phase as $\ket{\varphi}$.

$S$ is injective because we assume that $\ket{\varphi}$ is an injective MPS. Then the map $S^{-1}$ is well-defined on the support of $S$. In fact, we may identify the domain and support of $S$, and by applying an appropriate $U$ (replacing $S$ with $US$) assume that $S$ is hermitian and positive definite.

We now introduce a commuting Hamiltonian as follows. Consider $2N$ spins of dimension $\chi$ sitting on $N$ sites of a chain. So there are two spins on each site which are denoted by $l$ and $r$, and the Hilbert space corresponding to sites $j$ and $j+1$ is 
$$(\mathbb{C}^{\chi}_{lj}\otimes \mathbb{C}^{\chi}_{rj})\otimes (\mathbb{C}^{\chi}_{l(j+1)}\otimes \mathbb{C}^{\chi}_{r(j+1)}).$$
Define the local projection
$$P_{j,j+1}  = \openone_{lj}\otimes \openone_{rj}\otimes \openone_{l(j+1)}\otimes\openone_{r(j+1)}  -    \openone_{lj}\otimes   \ketbra{\Phi}{\Phi}_{rj, l(j+1)}   \otimes \openone_{r(j+1)},$$
and let $\widetilde H=\sum_j P_{j,j+1}$. $\widetilde H$ is commuting, translational invariant, and frustration free with the unique ground state $\ket{\Phi}\otimes \cdots \otimes \ket{\Phi}$. Moreover if we define
$$h_{j, j+1}  =  (S^{-1}_j\otimes S_{j+1}^{-1})  P_{j, j+1} (S_j^{-1}\otimes S_{j+1}^{-1}),$$ 
then $H=\sum_j h_{j, j+1}$ has a unique ground state which is $\ket{\varphi}$. As a result, the Hamiltonian $H$ with ground state $\ket{\varphi}$ can be turned into a commuting one using equation~\eqref{eq:localterm} with $X^{1/2} =S $. 

We conclude that MPS ground states and commuting Hamiltonians are in close relation via \eqref{eq:localterm}. The advantage of working with commuting Hamiltonians, however, is that verifying commutativity in general is much easier than checking whether the ground space is describable by MPSs.

We leave the problem of classification of phases of 1D systems for general Hamiltonians, without the commutativity assumption and that of \cite{Chen11, Schuch10}, for future works.

\myacknowledgments

The author is thankful to Norbert Schuch for explaining the difficulties
in removing the assumption of~\cite{Chen11,Schuch10} that the ground
space is describable by MPSs, and to Ramis Movassagh and unknown referees whose comments significantly improved the presentation of the paper.


{\small \vspace*{0.04in}
}{\small \par}

\end{document}